\documentclass[prl,amsmath,amssymb,showpacs,superscriptaddress,nofootinbib,twocolumn]{revtex4-1}
\usepackage{graphicx}
\usepackage{epstopdf}
\usepackage[mathlines]{lineno}
\usepackage{dcolumn}
\usepackage{color}
\usepackage{bm}
\usepackage{overpic}
\usepackage{rotating}
\usepackage{times}
\usepackage{multirow}
\usepackage{float}
\usepackage{mathrsfs}
\usepackage{booktabs}
\usepackage{float}
\usepackage[mathlines]{lineno}
\usepackage[T1]{fontenc}

\newcommand{\EE}{e^+e^-}

\newcommand{\psip}{\psi(3686)}
\newcommand{\psp}{\psip}

\newcommand{\OOb}{\Omega^-\bar{\Omega}^{+}}

\begin{document}
\hyphenpenalty=10000
\tolerance=1000

\title{\boldmath Model independent determination of the spin of the $\Omega^{-}$
and its polarization alignment in $\psip\to \OOb$}
\author{
M.~Ablikim$^{1}$, M.~N.~Achasov$^{10,c}$, P.~Adlarson$^{64}$,
S.~Ahmed$^{15}$, M.~Albrecht$^{4}$, A.~Amoroso$^{63A,63C}$,
Q.~An$^{60,48}$, ~Anita$^{21}$, Y.~Bai$^{47}$, O.~Bakina$^{29}$
% ...
    \\
        \vspace{0.2cm}
        (BESIII Collaboration)\\
        \vspace{0.2cm} {\it
    $^{1}$ Institute of High Energy Physics, Beijing 100049, People's Republic of China\\
% ...
$^{4}$ Bochum  Ruhr-University, D-44780 Bochum, Germany\\
% ...
$^{10}$ G.I. Budker Institute of Nuclear Physics SB RAS (BINP), Novosibirsk 630090, Russia\\
% ...
$^{15}$ Helmholtz Institute Mainz, Johann-Joachim-Becher-Weg 45, D-55099 Mainz, Germany\\
% ...
$^{21}$ Indian Institute of Technology Madras, Chennai 600036, India\\
% ...
$^{29}$ Joint Institute for Nuclear Research, 141980 Dubna, Moscow region, Russia\\
% ...
$^{47}$ Southeast University, Nanjing 211100, People's Republic of China\\$^{48}$ State Key Laboratory of Particle Detection and Electronics, Beijing 100049, Hefei 230026, People's Republic of China\\
% ...
$^{60}$ University of Science and Technology of China, Hefei 230026, People's Republic of China\\
% ...
$^{63}$ (A)University of Turin, I-10125, Turin, Italy; (B)University of Eastern Piedmont, I-15121, Alessandria, Italy; (C)INFN, I-10125, Turin, Italy\\$^{64}$ Uppsala University, Box 516, SE-75120 Uppsala, Sweden\\
% ...
$^{c}$ Also at the Novosibirsk State University, Novosibirsk, 630090, Russia\\
% ...
        }
}

\date{\today}

\begin{abstract}

We present an analysis of the process $\psip\to \OOb$
($\Omega^-\to K^-\Lambda$, $\bar{\Omega}^+\to K^+\bar{\Lambda}$,
$\Lambda\to p\pi^-$, $\bar{\Lambda}\to \bar{p}\pi^+$) based on a
data set of $448\times 10^6$ $\psip$ decays collected with the
BESIII detector at the BEPCII electron-positron collider. The
helicity amplitudes for the process $\psip\to \OOb$ and the decay
parameters of the subsequent decay $\Omega^-\to K^-\Lambda$
$(\bar{\Omega}^+\to K^+\bar{\Lambda})$ are measured for the first
time by a fit to the angular distribution of the complete decay
chain, and the spin of the $\Omega^-$ is determined to be $3/2$ for the first time since its discovery more than 50 years before.

\end{abstract}

\maketitle

The discovery of the $\Omega^-$~\cite{Barnes:1964pd} was a crucial
step in our understanding of the microcosmos. It was a great
triumph for the eight-fold way model of
baryons~\cite{GellMann:1962xb} and it led to the postulate of
color charge~\cite{Greenberg:1964pe}. A key feature of the eight-fold way
and the quark model is that the $\Omega^{-}$ spin is $J=3/2$,
a prediction that has never been unambiguously confirmed by experiment.
The current best determination of $J=3/2$ is based on an analysis~\cite{Aubert:2006dc}
that assumes the spins of both the $\Xi^{0}_{c}$ and the $\Omega^{0}_{c}$ are their quark model values of $J=1/2$.

One of the conceptually simplest processes in which a
baryon-antibaryon pair can be created is electron-positron
annihilation. In this Letter, two $\Omega^{-}$ spin hypotheses, $J=1/2$ or $J=3/2$,
are tested using the joint angular distribution of the
sequential decays of $e^{+}e^{-}\rightarrow\Omega^{-}\bar{\Omega}^{+}$ process.
For $J=1/2$ hypothesis, two form factors are needed in the production of
baryon-antibaryon pair in electron-positron annihilation, a clear vector polarization,
strongly dependent on the baryon direction is observed~\cite{Ablikim:2018zay,Ablikim:2019vaj}.
For $J=3/2$ hypothesis, the annihilation process involves four complex form
factors~\cite{Korner:1976hv}. In addition to vector polarization,
the spin-$3/2$ fermions can have quadrupole and octupole
polarization~\cite{Doncel:1972ez,Dubnickova:1992ii}. Polarization of the $\Omega^-$ can be
studied using the chain of weak decays $\Omega^-\to K^-\Lambda$
and $\Lambda\to p\pi^-$, where the first decay is described by the
ratio $\alpha_{\Omega^{-}}$ and the relative phase
$\phi_{\Omega^{-}}$ between the parity-conserving $P$-wave and
parity-violating $D$-wave ($S$-wave for $J=1/2$ hypothesis) decay amplitudes.
The decay parameters cannot be calculated reliably in
theory~\cite{Suzuki:1964,Hara:1966wlh,Finjord:1978ev} and only
$\alpha_{\Omega^{-}}$ has been previously
measured~\cite{Chen:2005aza,Lu:2005fc,Lu:2006bn}.

The resonance production process $\EE\to\psp\to \OOb$ was observed
by the CLEO-c experiment with $27\pm 5$ and $326\pm19$ events
using the double-tag and single-tag technique as described in
Refs.~\cite{Dobbs:2014ifa} and~\cite{Dobbs:2017hyd}, respectively.
With the world's largest $\psip$ data
sample of $(448.1\pm2.9)\times10^6$ $\psp$ events accumulated in
$\EE$ annihilation with BESIII detector~\cite{Ablikim:2009aa}, we
are able to select about 4000 $\psp\to \OOb$ events, and establish
for the first time that the $\Omega^-$ spin is $J=3/2$ and measure
$\Omega^-$ polarizations in the $\psp\to \OOb$ reaction, and
evidence for the dominance of the parity violating $D$-wave amplitude
in the weak decay $\Omega^{-}\to K^{-}\Lambda$.

\begin{figure}
\centering
   \includegraphics[width=0.45\textwidth]{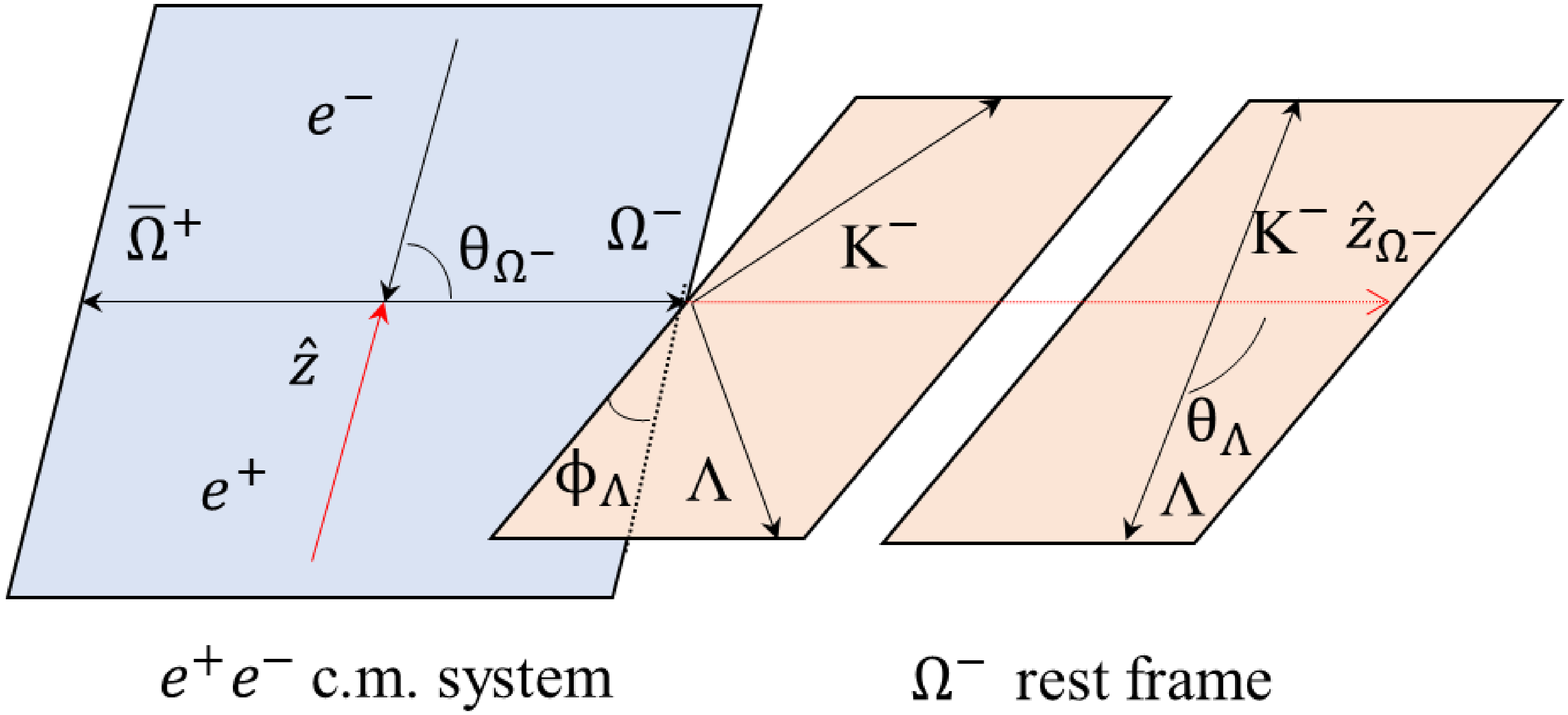}\\
   \vskip5pt
   \includegraphics[width=0.45\textwidth]{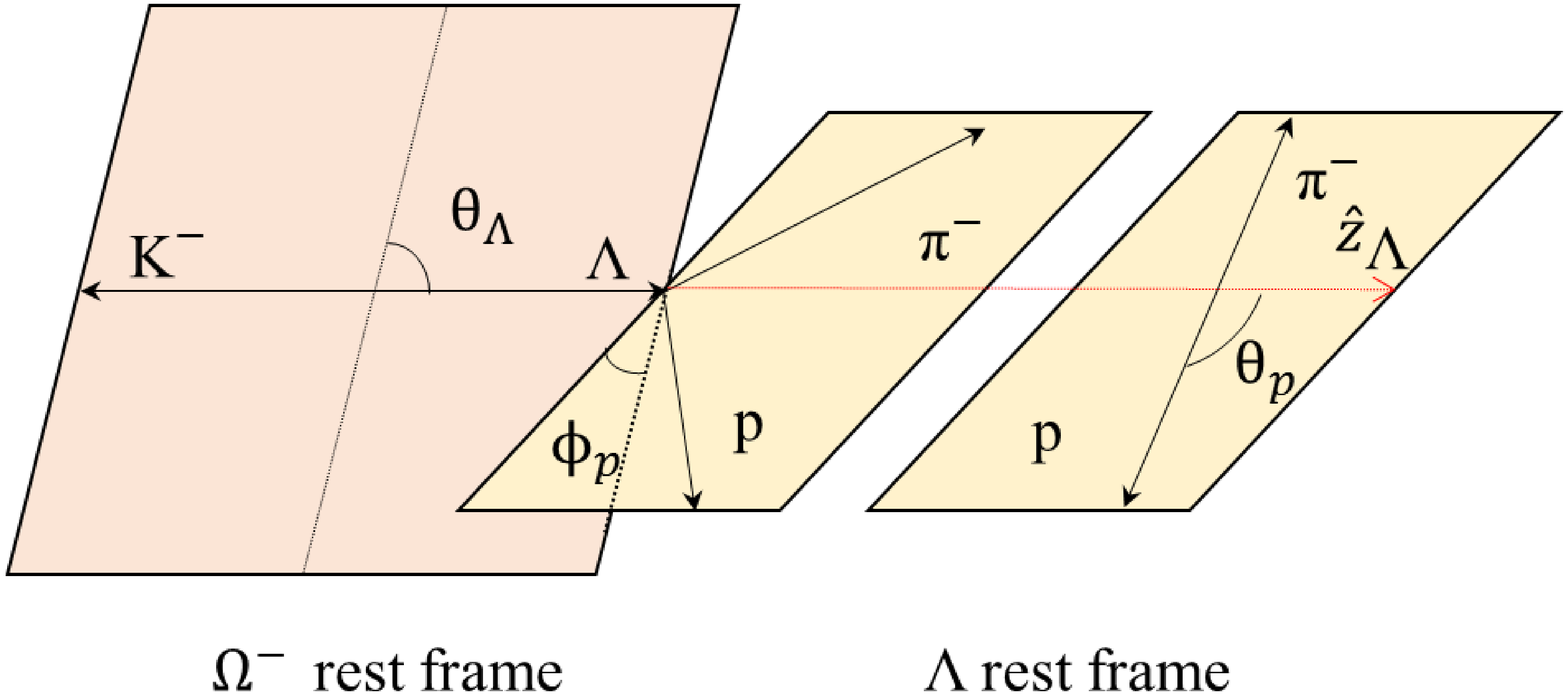}%
\caption{Definition of the helicity angles used in the analysis.
The helicity angles $\theta_{\Omega^{-}}$, $\theta_{\Lambda}$,
$\phi_{\Lambda}$, $\theta_{p}$ and $\phi_{p}$ are spherical
coordinates of the $\Omega^{-}$, $\Lambda$ and $p$ momenta in
three reference frames: the $e^+e^-$ c.m.\ system and the
$\Omega^{-}$ and $\Lambda$ rest frames, respectively. The
$\widehat{z}$-axis in the $e^+e^-$ c.m.\ system points along the
incoming positron and $\widehat{z}_{\Omega^{-}}$ is the $\Omega^-$
momentum direction. The polar axis direction in the $\Omega^-$
rest frame is $\widehat{z}_{\Omega^{-}}$ and
$\widehat{y}_{\Omega^{-}}$ is along $\widehat{z}\times
\widehat{z}_{\Omega^{-}}$, where $\widehat{z}_{\Lambda}$ is the
$\Lambda$ momentum direction. The polar axis direction in the
$\Lambda$ rest frame is $\widehat{z}_{\Lambda}$ and
$\widehat{y}_\Lambda$ is along $\widehat{z}_{\Omega^{-}}\times
\widehat{z}_{\Lambda}$.} \label{helicity_angle}
\end{figure}

For the $J=3/2$ hypothesis, in helicity formalism~\cite{Jacob:1959at,Berman:1965rc},
there are four helicity amplitudes in the production density matrix for
$\EE\to \psp\to \OOb$~\cite{Perotti:2018wxm}. We define the ratios
$A_{\frac{1}{2},\frac{1}{2}}/A_{\frac{1}{2},-\frac{1}{2}}=h_1e^{i\phi_1}$,
$A_{\frac{3}{2},\frac{1}{2}}/A_{\frac{1}{2},-\frac{1}{2}}=h_3e^{i\phi_3}$,
$A_{\frac{3}{2},\frac{3}{2}}/A_{\frac{1}{2},-\frac{1}{2}}=h_4e^{i\phi_4}$,
where, $h_i$ and $\phi_i~(i=1,3,4)$ are real numbers to be
determined from fits to data samples. The angular distribution is given
by the trace of the $\Omega^{-}$ spin density matrix~\cite{Perotti:2018wxm}:
$1+\alpha_{\psi(3686)}\cos^2\theta_{\Omega^{-}}$, where
$\alpha_{\psi(3686)} =
(1-2(|h_{1}|^{2}-|h_{3}|^{2}+|h_{4}|^{2}))/(1+2(|h_{1}|^{2}+|h_{3}|^{2}+|h_{4}|^{2}))$.
When considering the weak
decays $\Omega^-\to K^-\Lambda$ and $\Lambda\to p\pi^-$,
additional parameters $\alpha_{\Omega^-}$, $\alpha_{\Lambda}$ and
$\phi_{\Omega^-}$ describing the ratio and relative phase between
two helicity amplitudes are needed~\cite{Perotti:2018wxm}. The
joint angular distribution of $\theta_{\Omega^{-}}$,
$\theta_\Lambda$, $\phi_\Lambda$, $\theta_p$ and $\phi_p$ (see Fig.~\ref{helicity_angle}) is~\cite{Perotti:2018wxm}
\begin{equation}\label{angular_distribution}
\rho_{3/2}=\Sigma_{\mu=0}^{15}\Sigma_{\nu=0}^{3}r_{\mu}b_{\mu\nu}a_{\nu0}.
\end{equation}
For the $J=1/2$ hypothesis, the joint angular distribution is defined as~\cite{Perotti:2018wxm}:
\begin{equation}\label{1/2angular_distribution}
\rho_{1/2}=\Sigma_{\mu=0}^{3}\Sigma_{\nu=0}^{3}r_{\mu}a_{\mu\nu}a_{\nu0}.
\end{equation}
Here $r_{\mu}$, $b_{\mu\nu}$/$a_{\mu\nu}$, and $a_{\nu0}$ are defined in terms of
the helicity amplitudes~\cite{Perotti:2018wxm}. By fitting the joint angular distribution
of the selected events with Eqs.~(\ref{angular_distribution}) and (\ref{1/2angular_distribution}),
we can in principle obtain the
helicity amplitudes and $\Omega^-$/$\Lambda$ decay parameters.

To maximize the reconstruction efficiency, a single-tag method is implemented in
which only the $\Omega^-$ or the $\bar{\Omega}^+$ is reconstructed
via $\Omega^-\to K^-\Lambda\to K^-p\pi^-$ or $\bar{\Omega}^+\to
K^+\bar{\Lambda}\to K^+\bar{p}\pi^+$, and the $\bar{\Omega}^+$ or
$\Omega^-$ on the recoil side is inferred from the missing mass of
the reconstructed particles. The following
event selections are described for $\Omega^-\to K^-p\pi^-$ as an
example, the same selections are also applied for the
$\bar{\Omega}^+$ selection.

Charged tracks reconstructed from multilayer drift chamber (MDC)
hits are required to be within a polar-angle ($\theta$) range of
$|\cos\theta| < 0.93$. To determine the species of final-state
particles, specific energy loss ($dE/dx$) information is used to form particle
identification (PID) probabilities for pion, kaon, and proton
hypotheses. Charged particles are identified as the hypothesis
with the highest probability and only one $K^{-}$ and one proton are
required in each event. The rest of the negative charged tracks in an event
are assumed to be $\pi^{-}$. To avoid potentially large differences
between data and Monte Carlo (MC) simulation for very low momentum tracks, the
transverse momenta of the $p$, $K^-$, and $\pi^-$ tracks are
required to be larger than 0.2, 0.1, and 0.05~GeV/$c$,
respectively.

The $\Lambda\to p\pi^-$ candidates are reconstructed by applying a
vertex fit to the identified proton and a negatively charged pion
with an invariant mass ($M_{p\pi^{-}}$) in the mass window of
$[1.110,~1.122]~{\rm GeV}/c^2$. If more than one $\Lambda$
candidate is found, the one with $p\pi^{-}$ invariant mass closest
to the nominal $\Lambda$ mass~\cite{Tanabashi:2018oca} is kept.
The $\Lambda$ candidate is then combined with a $K^-$ track to
reconstruct the $\Omega^{-}$. A secondary vertex fit is applied to
$K^-\Lambda$ to improve the $\Omega^-$-mass resolution and to
suppress backgrounds. The invariant mass of $K^-\Lambda$ ($M_{K^-\Lambda}$)
is requirement in the mass window of $[1.663,~1.681]~{\rm GeV}/c^2$.
To obtain the antibaryon candidates $\bar{\Omega}^{+}$, we require
the recoiling mass of $K^-\Lambda$ ($M^{\rm recoil}_{K^-\Lambda}$) in
the mass window of $[1.640,~1.692]~{\rm GeV}/c^2$.
All the mass windows are determined by optimizing the figure of merit
$\frac{s}{\sqrt{s+b}}$ with $s$ being the number of signal events
expected in data and $b$ the number of the backgrounds in data
estimated by using a normalization factor of sideband regions and signal region.

\begin{figure}
\centering
\includegraphics[width=0.48\textwidth]{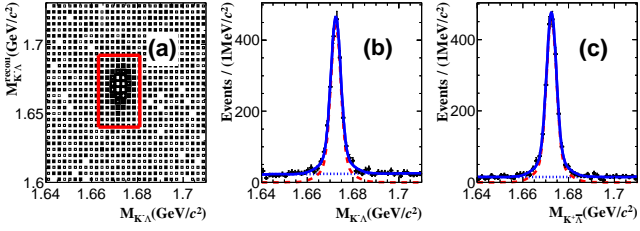}%
\caption{(a) Distribution of $M^{\rm recoil}_{K^-\Lambda}$ versus
$M_{K^-\Lambda}$ of the selected $K^-\Lambda$ candidates
in $\Omega^{-}$ reconstruction process. The red
solid box shows the signal region of $\psp\to \OOb$.
(b) Projection onto $M_{K^-\Lambda}$ for events with $M^{\rm
recoil}_{K^-\Lambda}$ in signal region. (c) Same as (b) but for
$K^+ \bar{\Lambda}$ tagged events. Dots with error bars are data,
the solid blue curves show the results of the fit, the red dashed
lines show the signal components of the fit, and the blue dotted
lines show the background components of the fit.} \label{scatter}
\end{figure}

The distribution of $M_{K^-\Lambda}$ vs. $M^{\rm recoil}_{K^-\Lambda}$ of the selected
$K^-\Lambda$ candidates is shown in Fig.~\ref{scatter}(a). A clear cluster of
events in the data sample corresponding to $\psp\to \OOb$ is
observed in the signal region of red box area.

An inclusive $\psp$ MC sample with $4\times 10^8$ $\psp$ events is
used to study the possible background sources~\cite{Zhou:2020ksj}
included in the simulation, and no peaking background is found.
The continuum production of $\Omega^{-}\bar\Omega^{+}$ is expected
to be very low and neglected. This is also checked with data
collected at 3.65~GeV with an integrated luminosity of
49~pb$^{-1}$ (about 7\% of the $\psi(3686)$ data sample), no
significant $\Omega^{-}\bar\Omega^{+}$ signal is observed.

Events in the signal region, shown in Fig.~\ref{scatter}(a), are
used to perform the angular distribution analysis. After applying
all the event selections, 2507~$\psp\to \OOb$ candidates are
selected by tagging the $\Omega^{-}$ (called the $\Omega^-$
sample), and 2238 candidates by tagging the $\bar{\Omega}^+$
(called the $\bar{\Omega}^+$ sample) by counting. The number of
non-$\Omega^-$ background events is estimated from the numbers of events in the
$\Omega^-$-mass sideband as $M_{K^-\Lambda}\in [1.644,~1.653]$ or
$[1.692,~1.701]~{\rm GeV}/c^2$. The $\Omega^-$ ($\bar{\Omega}^{+}$) sample
is estimated to contain $298\pm17$ ($189\pm14$) background events.

An unbinned maximum likelihood fit to the selected events is performed
to measure the free parameters in the angular distribution. The
likelihood function is defined as
\begin{equation}\label{likelihood_tot}
\mathcal{L}=\Pi_{j=1}^{N_t}W(\zeta_{j}|H) =
\Pi_{j=1}^{N_t}\frac{\rho(\zeta_{j}|H)\times
\epsilon(\zeta_{j})}{N(H)},
\end{equation}
where $j$ is the candidate event number, $\rho(\zeta_{j}|H)$ is
the angular distribution function for the cascade decay in
Eqs.~(\ref{angular_distribution}) and~(\ref{1/2angular_distribution}), $\zeta$ = $\{\theta_\Omega,
\theta_\Lambda, \phi_\Lambda, \theta_p, \phi_p\}$ are the angular
distribution variables, and $H$ contains the
parameters to be determined from the fit. $N_{t}$ is the number of
the selected events in the data samples. $N(H)$ is the normalization
factor calculated with the MC integration method, and
$\epsilon(\zeta_j)$ is the detection efficiency. Contributions
from the background events to the likelihood have been considered
by using events in the sideband regions of the $\Omega^{-}$.
The fit is performed by minimizing the objective function
   $S=-(\ln\mathcal{L}_{\rm data}-\ln\mathcal{L}_{\rm bg})$,
where $\mathcal{L}_{\rm data}$ is
the likelihood function of events selected in the signal region of $\Omega^{-}$ and $\bar{\Omega}^{+}$ samples,
and $\mathcal{L}_{\rm bg}$ is the likelihood function of
background events of these two single-tag samples estimated by the sideband method.

The decay parameters $\alpha_{\Lambda}$ and $\alpha_{\Omega^{-}}$ are
fixed to the PDG averages of previously measured
values~\cite{Ablikim:2018zay,Chen:2005aza,Lu:2005fc, Lu:2006bn}.
Assuming that there is no $CP$-violation in $\Omega^{-}$ and $\Lambda$
decays, $\alpha_\Lambda = -\alpha_{\bar{\Lambda}} = 0.753
\pm 0.007$ and $\alpha_{\Omega^{-}} = -\alpha_{\bar{\Omega}^{+}} =
0.0154 \pm 0.0017$~\cite{Explanation}. A simultaneous fit is performed to the
$\Omega^-$ and $\bar{\Omega}^+$ events selected from data in which the constraint $\phi_{\Omega^{-}} =
-\phi_{\bar{\Omega}^{+}}$ is applied. The change of $2S$ of the fit assuming $J=1/2$ and that of
a linear combination of $J=1/2$ and $J=3/2$ is $-232$ with 8 more free parameters,
so we determine the significance of the $J=3/2$ hypothesis over the
$J=1/2$ to be larger than $14\sigma$, thus determine the spin of $\Omega^-$ as 3/2 unambiguously.
For the fit with $J=3/2$, we find two solutions with identical fit quality, as
shown in Table~\ref{simultaneous_result}. Tests with large MC sample confirm the existence of two solutions in such fits although its origin is not obvious in the expression of the decay amplitude. The statistical and systematic covariance matrices for the two solutions are supplied in the Supplemental Material.

\begin{table}[htbp]
 \centering
\caption{Two sets of fit values of the helicity parameters in
$\psp\to\OOb$ decays of spin-3/2 hypothesis. The first uncertainties are statistical, and
the second ones systematic.}\label{simultaneous_result}
   \begin{tabular}{c|c|c}
   \hline\hline
   parameter & solution I & solution II  \\\hline
   $h_1$           & 0.30$\pm$0.11$\pm$0.04   & 0.31$\pm$0.10$\pm$0.04\\
   $\phi_1$        & 0.69$\pm$0.41$\pm$0.13   & 2.38$\pm$0.37$\pm$0.13\\
   $h_3$           & 0.26$\pm$0.05$\pm$0.02   & 0.27$\pm$0.05$\pm$0.01\\
   $\phi_3$        & 2.60$\pm$0.16$\pm$0.08   & 2.57$\pm$0.16$\pm$0.04\\
   $h_4$           & 0.51$\pm$0.03$\pm$0.01   & 0.51$\pm$0.03$\pm$0.01\\
   $\phi_4$        & 0.34$\pm$0.80$\pm$0.31   & 1.37$\pm$0.68$\pm$0.16\\
   $\phi_{\Omega}$ & 4.29$\pm$0.45$\pm$0.23   & 4.15$\pm$0.44$\pm$0.16\\
   \hline \hline
   \end{tabular}
\end{table}

\begin{figure*}[htbp]
  \centering
    \includegraphics[width=0.24\textwidth]{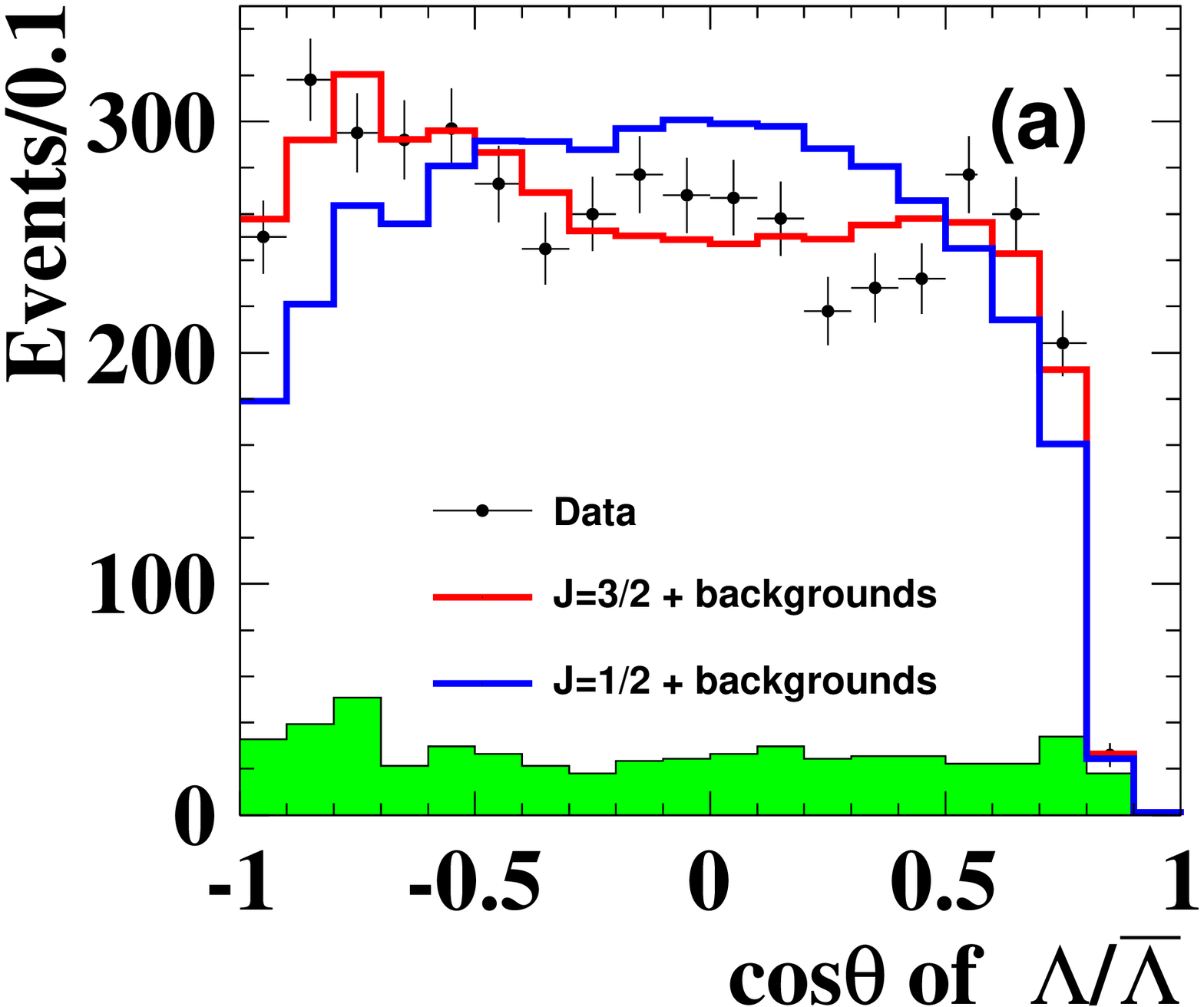}%
    \includegraphics[width=0.24\textwidth]{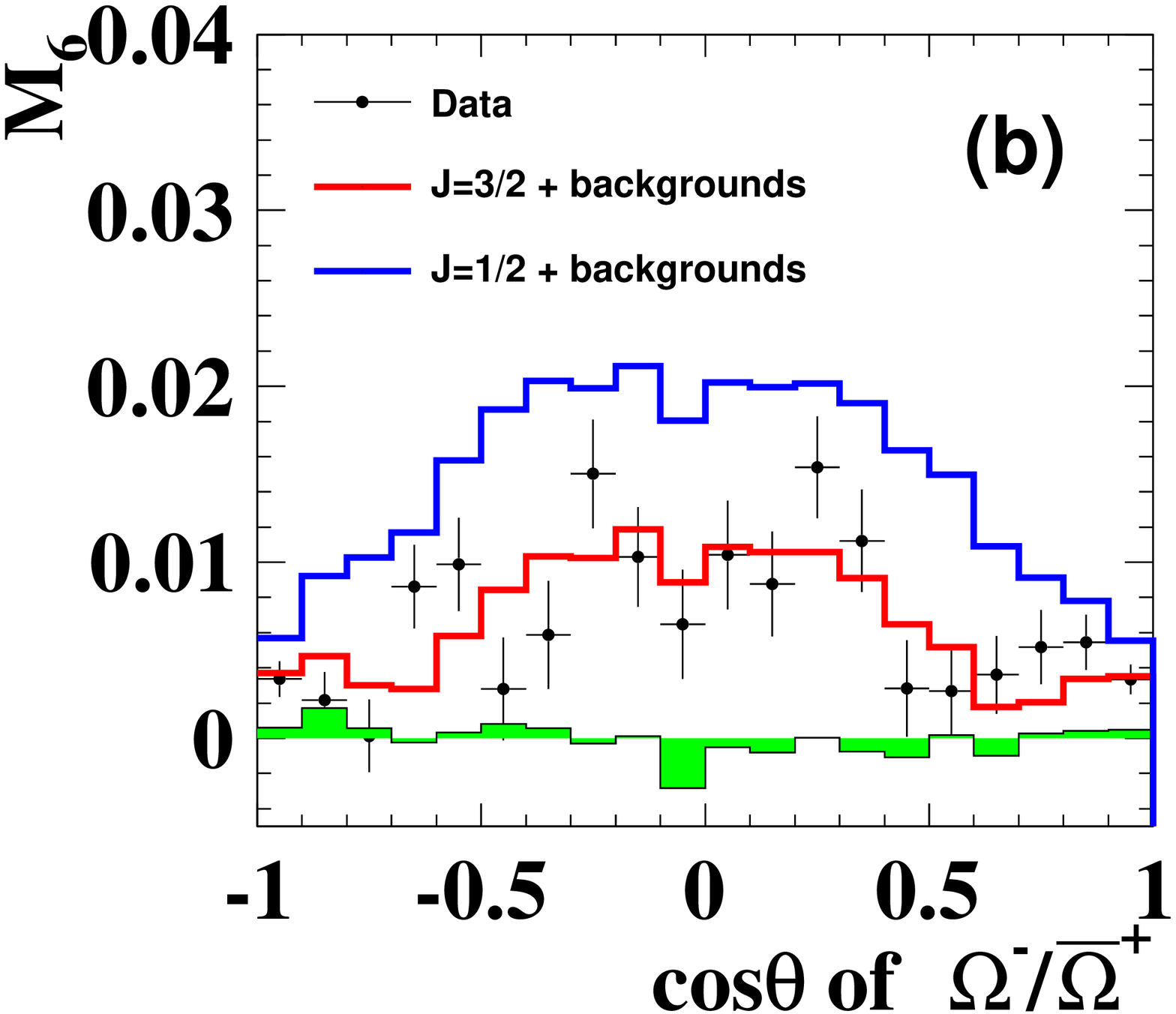} %
    \includegraphics[width=0.24\textwidth]{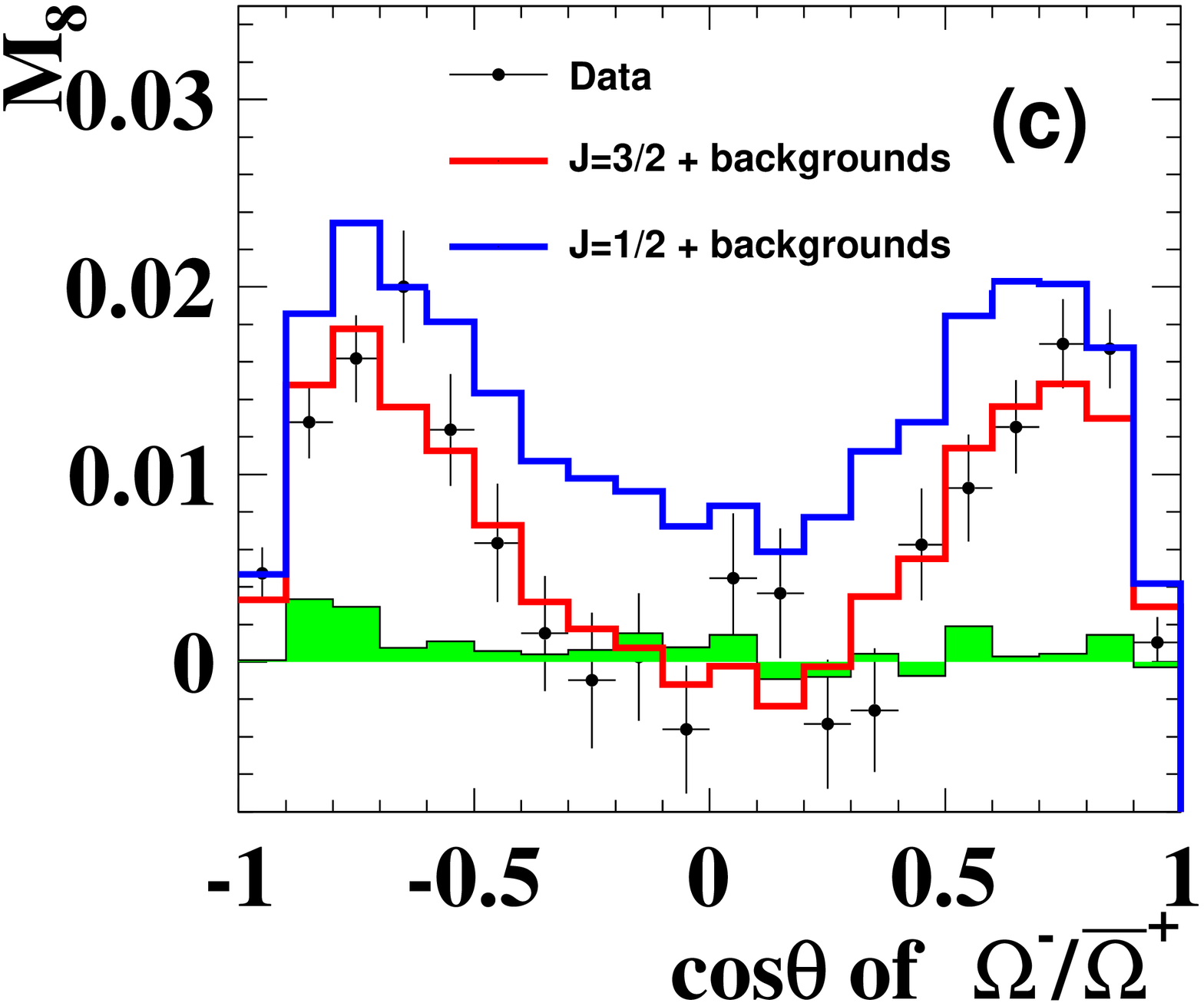}%
     \includegraphics[width=0.24\textwidth]{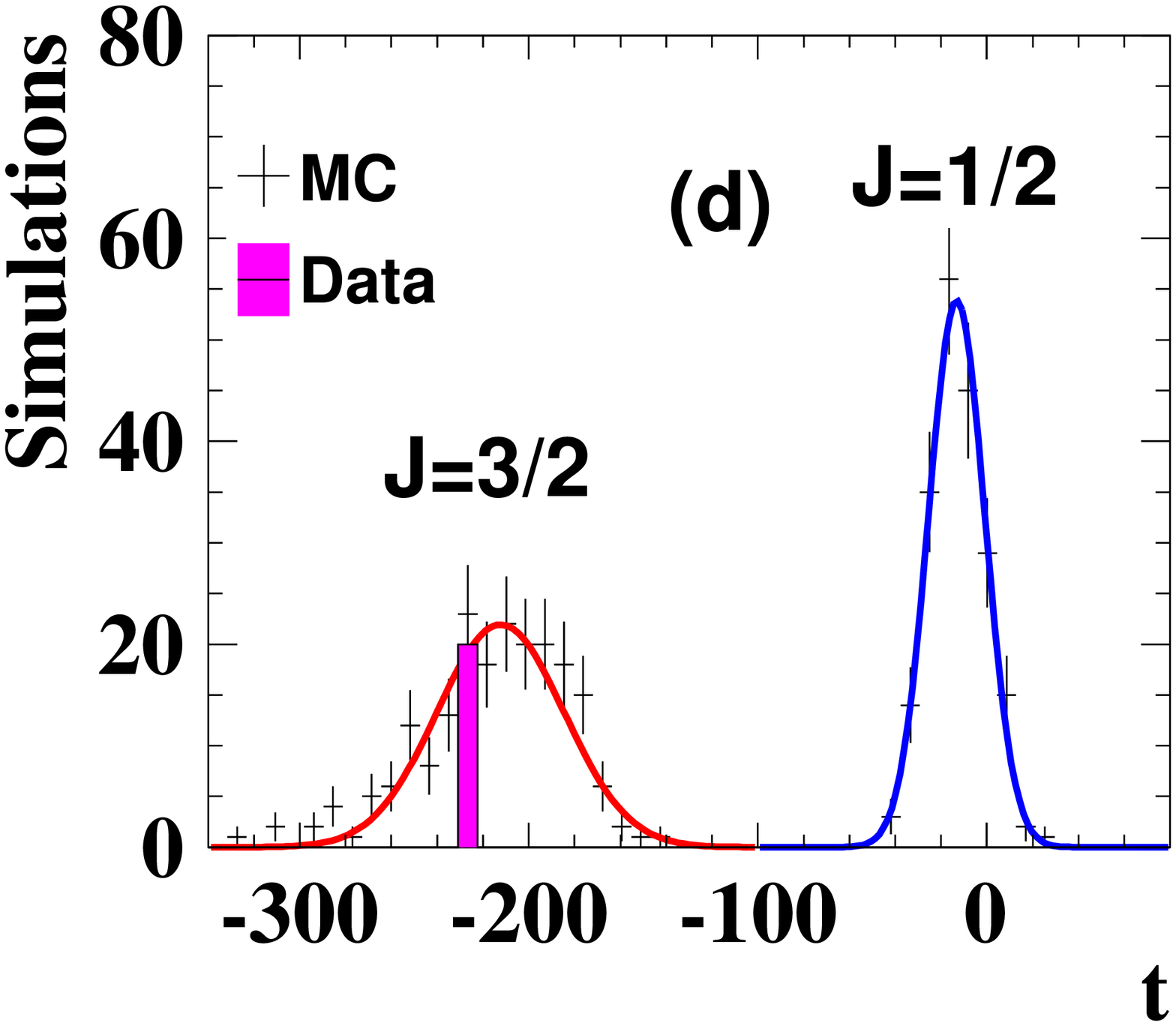}\\%
   \caption{(a) The $\cos\theta_{\Lambda/\bar{\Lambda}}$ distributions of data
  (dots with error bars) and fits with $J=3/2$ (red histogram) and $J=1/2$ (blue histogram) hypotheses; (b) and (c) are the $M_{6}$ and $M_{8}$ distributions of data and fit results; and (d) distribution of the test statistic $t=S^{J=1/2}-S^{J=3/2}$ for a series of MC simulations performed under the $J=1/2$ (right peak) and $J=3/2$ (left peak) hypotheses. The lines represent Gaussian fits to the simulated data points. The $t$ value obtained from experimental data is indicated by the vertical bar.}\label{angular}
\end{figure*}

The signal MC events generated with PHSP are weighted with matrix
elements calculated with the parameters obtained from the fits and
the weighted MC sample predictions are compared with data in five
distributions of the helicity angle, with the background
contributions estimated from the $\Omega^-$ sideband regions indicated as green histogram.
We observe that the fit with $\Omega^-$ spin $J=3/2$ describes data very
well while $J=1/2$ fails to describe data, as shown
in Fig.~\ref{angular}(a) for $\cos\theta_{\Lambda/\bar{\Lambda}}$, which
has the most prominent difference. The moments $M_{6}$ and $M_{8}$ defined as
$M_{\mu}=\frac{1}{N}\sum_{j=0}^{N}\sum_{k=0}^{3}b_{\mu,\kappa}a_{\kappa,0}$
are compared between data and those two weighted MC samples,
as shown in Figs.~\ref{angular} (b) and~(c). Here $N$ is the number of events in the data or MC samples. Clear preference of $J=3/2$ over
$J=1/2$ is observed. Since the two sets of solutions describe
the data equally well, we only show the angular distributions for
one set of them.

The likelihood ratio $t=2(S^{J=1/2}-S^{J=3/2})$ is used as
a test variable to discriminate between the $J=3/2$ and $J=1/2$ hypotheses~\cite{Aaij:2015eva}.
The MC sample for each hypothesis is generated according to its joint angular distribution,
propagated through the detector model and subjected to the same event selection criteria
as the experimental data. Each MC subset has the same size as the real data samples
and is assumed to have the same amount of background. The test statistic $t$ distribution
is shown in Fig.~\ref{angular}(d). The simulations for the right peak were performed under
the $J=1/2$ hypothesis, while those in the left peak correspond to the $J=3/2$ hypothesis.
It is clear that the $t$ distributions of the two hypotheses are well separated. Since
the $t$-value from the data lies well within the left peak, our data favour
the $J=3/2$ hypothesis.

The following systematic uncertainties are considered for the
angular distribution measurement. The tracking and PID
efficiencies are studied with control samples of $J/\psi\to
p\bar{p}\pi^+\pi^-$, $J/\psi\to \Lambda\bar{\Lambda}$ ($\Lambda\to
p\pi^-$, $\bar{\Lambda}\to \bar{p}\pi^+$), $J/\psi\to
$$K_S K^-\pi^++c.c.$, and $J/\psi\to
$$pK^-\bar{\Lambda}+c.c.$, and the polar angle and transverse
momentum ($p_t$) dependent efficiencies are measured.
Subsequently, the efficiency of MC events is corrected by the
two-dimensional efficiency scale factors and the uncertainty is
estimated by varying the efficiency scale factors by one standard
deviation for each $p_{t}$ vs. $\cos\theta$ bin.
The differences between the new fit results and
the nominal ones are taken as the systematic uncertainties. The
uncertainty due to the background estimation is estimated by
changing the $\Omega^{-}$-sideband regions from $[1.644,~1.653]$
and $[1.692,~1.701]~{\rm GeV}/c^2$ to $[1.643,~1.653]$ and
$[1.692,~1.702]~{\rm GeV}/c^2$. The differences between fit
results with and without changing sideband regions are taken as
the systematic uncertainties. The uncertainties arising from the
values of the fixed parameters $\alpha_{\Omega^{-}}$ and
$\alpha_{\Lambda}$ are estimated by changing these two parameters
by one standard deviation separately, then comparing the refitted
parameters with the original results. We find that the uncertainty
of $\alpha_{\Omega^{-}}$ can be neglected. All of the above
contributions are added in quadrature to obtain the total systematic
uncertainties as shown in Table~\ref{systematic_uncertanty}.

\begin{table}
 \centering
\caption{Summary of the systematic uncertainties for the decay
parameters in solution I (solution II) of $\psp\to \OOb$.}
\label{systematic_uncertanty}
   \begin{tabular}{c|c|c|c|c}
   \hline
   \hline
     & track/PID &background
   & $\alpha_{\Lambda}$  &total\\\hline
   $\Delta h_1$     &0.04(0.04)   & 0(0.01) &0(0)& 0.04(0.04)\\
   $\Delta\phi_1$  &0.13(0.12) & 0.02(0.04)   &0.01(0.01)  & 0.13(0.13)   \\
   $\Delta h_3$     &0.01(0.01)  & 0(0)   &0.02(0)  & 0.02(0.01) \\
   $\Delta\phi_3$  &0.03(0.03)   & 0.07(0.02)   &0(0.01)  & 0.08(0.04) \\
   $\Delta h_4$     &0.01(0.01)  & 0(0.01)   &0(0)  & 0.01(0.01) \\
   $\Delta\phi_4$  &0.28(0.11)  & 0.13(0.12)   &0.02(0)  & 0.31(0.16) \\
   $\Delta\phi_{\Omega}$  &0.16(0.16)  & 0.17(0.03)   &0(0.01)  & 0.23(0.16) \\
   \hline \hline
   \end{tabular}
   \end{table}

From Table~\ref{simultaneous_result} we find that the magnitudes
of the amplitudes are about the same in the two solutions while
the phases $\phi_{1}$ and $\phi_{4}$ can be very different.
All the $h_i$ values are less than one, which means that the amplitude
$A_{\frac{1}{2},-\frac{1}{2}}$ dominates the decay process.
The value of $\phi_{\Omega^-}$ provides information on whether the
process is $P$-wave dominant ($\phi_{\Omega^-}=0$) or $D$-wave
dominant ($\phi_{\Omega^-}=\pi$). By comparing the
maximum-likelihood values between the fit with $\phi_{\Omega^-}$
fixed to zero or $\pi$ and the nominal fit, we find that the
significance for non-zero $\phi_{\Omega^-}$ is $3.7\sigma$ and
that for a non-$\pi$ $\phi_{\Omega^-}$ is $1.5\sigma$. Thus,
$\phi_{\Omega^-}$ favours the $D$-wave dominant case, which differs
from the theoretical predictions of $P$-wave
dominance~\cite{Tandean:2004mv}. The ratio of $D$- to $P$-wave can
be calculated as $|A_{D}|^{2}/|A_{P}|^{2} = 2.4\pm2.0$
(solution~I), and $|A_{D}|^{2}/|A_{P}|^{2} = 3.3\pm2.9$
(solution~II), where the uncertainty is the sum in quadrature of
the statistical and systematic uncertainties.
Allowing $\alpha_{\Omega^{-}}$
to be determined by the fit, we obtain $\alpha_{\Omega^{-}}=-0.04\pm 0.03$,
which does not contradict with the quoted result from previous
experiments but with poorer
precision~\cite{Chen:2005aza,Lu:2005fc, Lu:2006bn}.


To conclude, based on $448\times 10^6$ $\psp$ events, we observe
4035 $\pm$ 76~$\psp\to \OOb$ signal events. We conduct the first
study of the angular distribution of the three-stage decay,
and found that the hypothesis of $\Omega^-$ has a spin of 3/2
is preferred over a spin of 1/2 with a significance of
more than 14$\sigma$, and establishes the spin of the
$\Omega^{-}$ to be 3/2 for the first time that is independent of any model-based assumptions.
The helicity amplitudes of $\psp\to \OOb$ and the decay parameter of
$\Omega^{-}\to K^{-}\Lambda$, $\phi_{\Omega^{-}}$, are also measured for the first
time. With the helicity amplitudes measured in
Table~\ref{simultaneous_result}, $\alpha_{\psp}=0.24\pm 0.10$,
where the uncertainty is the sum in quadrature of the statistical
and systematic uncertainties.

\begin{figure}
\centering
    \includegraphics[width=0.3\textwidth]{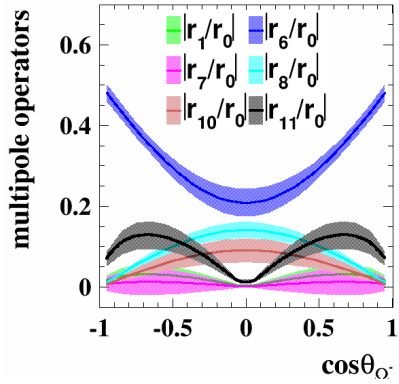}
\caption{The $\cos\theta_{\Omega^-}$ dependence of the multipolar polarization operators.
The solid lines represent the central values, and the shaded areas
represent $\pm$ one standard deviation.}
\label{polarization_Omega}
\end{figure}

With the helicity amplitudes measured in Table~\ref{simultaneous_result},
we calculate the $\cos\theta_{\Omega^{-}}$ dependence of the multipolar
polarization operators as shown in Fig.~\ref{polarization_Omega}.
The uncertainties (statistical and systematic) are calculated
using the covariance matrix of the fitted $h_{i}$ and $\phi_{i}$.
For the process of $e^{+}e^{-}\to \psi(3686)\to \Omega^{-}\bar{\Omega}^{+}$,
$\Omega^{-}$ particles not only have vector polarization ($r_{1}$), but also have
quadrupole ($r_{6}$, $r_{7}$, $r_{8}$) and octupole ($r_{10}$, $r_{11}$)
polarization contributions~\cite{Doncel:1972ez,Dubnickova:1992ii}.

As a byproduct, with the same data sample, the branching fraction
for $\psi(3686)\rightarrow\Omega^{-}\bar{\Omega^{+}}$ is measured
as $(5.85\pm 0.12\pm 0.25)\times 10^{-5}$, where the first
uncertainty is statistical systematic, and the second is systematic~\cite{supple}. This
result agrees with previous
measurements~\cite{Dobbs:2014ifa,Dobbs:2017hyd} with improved
precision.

The BESIII collaboration thanks the staff of BEPCII and the IHEP
computing center for their strong support. The authors would like
to thank Prof.~Zuotang Liang, Prof. Yukun Song, Prof. Xiaogang He,
and Dr. Jusak Tandean for useful discussions. This work is
supported in part by National Key Basic Research Program of China
under Contract No. 2020YFA0406300; National Natural Science
Foundation of China (NSFC) under Contracts Nos. 11625523,
11635010, 11735014, 11822506, 11835012, 11935015, 11935016,
11935018, 11961141012; the Chinese Academy of Sciences (CAS)
Large-Scale Scientific Facility Program; Joint Large-Scale
Scientific Facility Funds of the NSFC and CAS under Contracts Nos.
U1732263, U1832207; CAS Key Research Program of Frontier Sciences
under Contracts Nos. QYZDJ-SSW-SLH003, QYZDJ-SSW-SLH040; 100
Talents Program of CAS; INPAC and Shanghai Key Laboratory for
Particle Physics and Cosmology; ERC under Contract No. 758462;
German Research Foundation DFG under Contracts Nos. 443159800,
Collaborative Research Center CRC 1044, FOR 2359, GRK 214;
Istituto Nazionale di Fisica Nucleare, Italy; Ministry of
Development of Turkey under Contract No. DPT2006K-120470; National
Science and Technology fund; Olle Engkvist Foundation under
Contract No. 200-0605; STFC (United Kingdom); The Knut and Alice
Wallenberg Foundation (Sweden) under Contract No. 2016.0157; The
Royal Society, UK under Contracts Nos. DH140054, DH160214; The
Swedish Research Council; U. S. Department of Energy under
Contracts Nos. DE-FG02-05ER41374, DE-SC-0012069.


\end{document}